\documentstyle[seceq]{ptptex}

\newcommand{\bom}[1]{\mbox{\boldmath{$#1$}}}
\newcommand{\boml}[1]{\mbox{\boldmath{$#1$}}^{||}}
\newcommand{\bomt}[1]{\mbox{\boldmath{$#1$}}^{\bot}}

\newcommand{\bomlo}[2]{\mbox{\boldmath{$#1$}}^{||}_{(#2)}}
\newcommand{\bomto}[2]{\mbox{\boldmath{$#1$}}^{\bot}_{(#2)}}

\newcommand{\lp}[1]{#1^{||}}
\newcommand{\tp}[1]{#1^{\bot}}

\newcommand{\lpo}[3]{#1^{||}_{(#2)#3}}
\newcommand{\tpo}[3]{#1^{\bot}_{(#2)#3}}
\newcommand{\dlpo}[3]{\dot{#1}^{||}_{(#2)#3}}

\newcommand{\nx}{\nabla_{x}}
\newcommand{\nq}{\nabla_{q}}

\title{
Non-linear vorticity-density coupling in Lagrangian dynamics
}

\author{
Makoto {\sc Sasaki} and Masumi {\sc Kasai}
}

\inst{
Department of Physics, Hirosaki University, Hirosaki 036
}

\notypesetlogo

\abst{
We present a general Lagrangian formalism
that allows the treatment of vorticity.
We give solutions for the rotational perturbations up to the third-order
in a flat background universe.
We show how the primordial vorticity affects the evolution 
of the density fluctuation in high-density regions.
}

\begin{document}
\maketitle

\section{Introduction}
\label{sec:introduction}

The study of the galaxy formation based on gravitational instability
has not revealed the origin of the vorticity of galaxies.
The possibility that galactic spins are the relics of
primordial vorticity has not been taken into account.
Because the standard linear Eulerian theory 
(see, for example, Peebles' book
\cite{Peebles}
for the detail of the Eulerian theory)
shows that the vorticity
decays at late time and does not couple to the density enhancement.
For this reason,
the methods that generate the vorticity posteriorly are investigated.
Doroshkevich
\cite{Doroshkevich}
explained the acquisition of the vorticity
by the formation of shock fronts in the pancake.
The explanation by tidal torques was made by Hoyle.
\cite{Hoyle} \ 

However,
can we actually ignore the effects of the primordial vorticity
in the galaxy formation process?
To answer this question,
we must reconsider the validity of the Eulerian theory.
In the Eulerian treatment,
density fluctuation is perturbative quantity,
so it must be small.
Therefore,
the Eulerian approach is limited to the description
of weak density fluctuation.
For this reason,
to discuss the behavior of the vorticity in the high-density regions,
we must prepare the theory beyond the Eulerian theory.
The Lagrangian theory improves such shortcomings of the Eulerian theory.
The most advantageous point of the Lagrangian theory
is that
the density fluctuation is non-perturbative quantity.
Therefore,
it is unnecessary that the density fluctuation is expanded
in a perturbation series
and that we impose the condition of the smallness of the density fluctuation.
By using the property of the Lagrangian theory,
we can investigate the behavior of the primordial vorticity.
Buchert
\cite{Buchert1} \ 
solved the first-order Lagrangian perturbative equations
and showed that the primordial vorticity was amplified
in proportion to the enhancement of the density fluctuation
by deriving the relation between the vorticity and the density fluctuation.
Barrow and Saich
\cite{Barrow}
solved the vorticity equation on the inhomogeneous background
and pointed out that
the growing directions of the vorticity 
were lying in the plane of the pancake.
These analyses based on the Lagrangian theory
shows that the primordial vorticity is not negligible 
in the high-density regions such as the pancake.

In this paper,
for the purpose of preparing a tool to treat
the behavior of the vorticity in the high-density regions,
we extend the previous works
and give solutions including the vorticity
for the Lagrangian perturbative equations
up to the third-order.
Furthermore,
we obtain the relation between the vorticity and the density fluctuation. 
The reason we proceed to the calculation up to the third-order
is that the third-order solutions will cover the main effects
since the density fluctuation is described
by the determinant of a $3\times 3$ matrix. 
In the Lagrangian approach,
we assume that 
background is Friedmann-Lema\^{\i}tre-Robertson-Walker (FLRW) universe
with no cosmological constant and curvature.
There is no definitive observational evidence that 
our universe is actually flat.
We apply this assumption for simplifying calculations.

This paper is organized as follows.
In \S\ref{sec:Basic equations in Lagrangian form},
we summarize the basic formulae of the Lagrangian theory.
In \S\ref{sec:Lagrangian perturbative approximation},
we present the perturbative approach
to the Lagrangian equations
including the treatment of the vorticity. 
In \S\ref{Summary and conclusions},
we summarize the main results of this paper.


\section{Basic equations in Lagrangian form}
\label{sec:Basic equations in Lagrangian form}

In this section,
we give a summary of the Lagrangian theory.
We particularly focus on 
the treatment of the vorticity in the Lagrangian theory.
The detail of the Lagrangian theory is found,
for example,
in Ref.~\citen{Ehlers}.


\subsection{Density fluctuation in Lagrangian form}
\label{sec:Density fluctuation in Lagrangian form}

In this subsection,
we focus on the expression of the density fluctuation in Lagrangian form.

In the Eulerian perturbation theory,
the density fluctuation $\delta$ is a perturbative quantity.
On the other hand,
in the Lagrangian form,
the density fluctuation
is given in the formally exact form as follows:
\begin{equation}
\label{def:density}
  \delta=J^{-1}-1,
  \mbox{ }
  J:=\det\left(\frac{\partial x_{i}}{\partial q_{j}}
         \right),
\end{equation}
where $\bom{x}$ are expanding coordinates in the Eulerian space
and $\bom{q}$ are the Lagrangian coordinates
which are defined by the initial Eulerian position,
namely,
$\bom{q}:=\bom{x}(t=t_{0})$.
(The physical distance is obtained according to the law
 $\bom{r}=a(t)\bom{x}$,
 $a(t)$ being the scale factor.)
The determinant $J$ of the Jacobian of the transformation
from $\bom{x}$ to $\bom{q}$
is basic quantity to represent the density fluctuation.
Since $J$ is defined by the Lagrangian partial derivative,
we must evaluate $\bom{x}$ in the Lagrangian space:
$\bom{x}=\bom{x}(\bom{q},t)$.
Therefore,
it is necessary 
to clarify the relation 
between the Eulerian coordinates and the Lagrangian ones.
The displacement vector is the quantity that relates $\bom{x}$ with $\bom{q}$.
This quantity represents time evolution from the initial position $\bom{q}$.
By using the displacement vector $\bom{A}(\bom{q},t)$,
we can write the relation between $\bom{x}$ and $\bom{q}$ as
$\bom{x}=\bom{q}+\bom{A}(\bom{q},t)$.
If we obtain the concrete form of $\bom{A}$,
the density fluctuation is calculated from Eq.~(\ref{def:density}).
Since Eq.~(\ref{def:density}) is derived from mass conservation
between the Eulerian and the Lagrangian space,
this equation is hold irrespective of the existence of the vorticity.

Now we shall consider 
how the effects of the vorticity on the density fluctuation are reflected.
A conventional way
to investigate the effects of the vorticity
is that we split $\bom{A}$ into the longitudinal mode and the transverse one
with respect to the the Eulerian coordinates
and then focus on the behavior of the transverse mode.
However,
since $\bom{A}$ is evaluated in the Lagrangian coordinates,
we shall perform this decomposition
with respect to the Lagrangian coordinates:
\begin{equation}
  \bom{A}=\boml{A}+\bomt{A},
  \mbox{ }
  \nq\times\boml{A}=0,
  \mbox{ }
  \nq\cdot\bomt{A}=0,
\end{equation}
where the superscript $\Vert$ and $\bot$ denote 
the longitudinal and the transverse mode 
with respect to the Lagrangian coordinates,
respectively.
Although the quantity $\bomt{A}$ does not correspond with the vorticity,
we shall express the density fluctuation by using $\boml{A}$ and $\bomt{A}$.
The method to express the vorticity in the Lagrangian space 
will be presented in next subsection.
From the definition of $J$ in Eq.~(\ref{def:density}),
we obtain
\begin{eqnarray}
\label{def:Jacobian}
  J
  &=&
  1+\nq\cdot\boml{A}\nonumber\\
  &&\mbox{ }
  +\frac{1}{2}\left[\left(\nq\cdot\boml{A}\right)^2
                    -\lp{A}_{i,j}\lp{A}_{j,i}
              \right]
  -\lp{A}_{i,j}\tp{A}_{j,i}-\frac{1}{2}\tp{A}_{i,j}\tp{A}_{j,i}
  \nonumber\\
  &&\mbox{ }
  +\det\left(\lp{A}_{i,j}+\tp{A}_{i,j}\right),
\end{eqnarray}
where 
\begin{eqnarray}
  \det(\lp{A}_{i,j}+\tp{A}_{i,j})
  &=&
  \frac{1}{3}\lp{A}_{i,j}\lp{A}_{j,k}\lp{A}_{k,i}
  -\frac{1}{2}\lp{A}_{i,j}\lp{A}_{j,i}\lp{A}_{l,l}
  +\frac{1}{6}\left(\lp{A}_{i,i}\right)^3\nonumber\\
  &&\mbox{ }
  +\lp{A}_{i,j}\tp{A}_{j,k}\lp{A}_{k,i}
  -\lp{A}_{i,j}\tp{A}_{j,i}\lp{A}_{l,l}\nonumber\\
  &&\mbox{ }
  +\tp{A}_{i,j}\tp{A}_{j,k}\lp{A}_{k,i}
  -\frac{1}{2}\tp{A}_{i,j}\tp{A}_{j,i}\lp{A}_{l,l}\nonumber\\
  &&\mbox{ }
  +\frac{1}{3}\tp{A}_{i,j}\tp{A}_{j,k}\tp{A}_{k,i}
\end{eqnarray}
and the comma denotes the partial derivative 
with respect to the Lagrangian coordinates.
By considering the transverse mode $\bomt{A}$,
we can investigate
how the vorticity contributes to the density enhancement.

In next subsection,
we derive the Lagrangian equations for $\boml{A}$ and $\bomt{A}$
which are necessary for the calculation of the density fluctuation.


\subsection{Equations for longitudinal and transverse mode
            in Lagrangian coordinates}

As mentioned in previous subsection,
we derive the Lagrangian equations for $\boml{A}$ and $\bomt{A}$ here.
The derivation of the equations is based on the methods
by Buchert and G\"{o}tz,
\cite{Gotz}
and Kasai
\cite{Kasai}.
Furthermore,
we also present the expression of the vorticity in the Lagrangian space.

First,
we shall derive the Lagrangian equations.
We denote 
the density of dust, the background density, the peculiar velocity 
and the peculiar gravitational acceleration
by $\rho$, $\rho_{b}$, $\bom{v}$ and $\bom{g}$, respectively.
The peculiar velocity 
given by the form $\bom{v}=a\dot{\bom{x}}$
represents the deviation from the uniform Hubble flow.
Since we assume the FLRW universe,
the scale factor $a$ is proportional to $t^{2/3}$.
We consider that $a=t^{2/3}$.
The basic equations in Newtonian cosmology that these quantities obey are
\begin{subequations}
\label{sys:basic} 
  \begin{equation}
  \label{eq:basic1}
    \frac{\partial\rho}{\partial t}+3\frac{\dot{a}}{a}\rho
    +\frac{1}{a}\nabla_{x}\cdot(\rho\bom{v})=0 ,
  \end{equation}
  \begin{equation}
  \label{eq:basic2}
    \frac{\partial\bom{v}}{\partial t}
    +\frac{\dot{a}}{a}\bom{v}
    +\frac{1}{a}(\bom{v}\cdot\nabla_{x})\bom{v}
    =\bom{g},
  \end{equation}
  \begin{equation}
  \label{eq:basic3}
    \nabla_{x}\times\bom{g}=0,
  \end{equation}
  \begin{equation}
  \label{eq:basic4}
    \nabla_{x}\cdot\bom{g}
    =-4\pi Ga(\rho-\rho_{b}),
  \end{equation}
\end{subequations}
where $\nx$ is the Eulerian nabla operator.
We obtain
from Eqs.~(\ref{eq:basic1}) and (\ref{eq:basic4})
\begin{subequations}
\label{sys:basic_sum}
  \begin{equation}
  \label{eq:basic1+basic4}
    \nabla_{x}\cdot
    \left(
      \frac{\partial \bom{g}}{\partial t}
      +2\frac{\dot{a}}{a}\bom{g}
      -4\pi G \rho\bom{v}
    \right)
    =0,
  \end{equation}
and from Eqs.~(\ref{eq:basic2}) and (\ref{eq:basic3})
  \begin{equation}
  \label{eq:basic2+basic3}
    \nx\times
    \left(
      \frac{\partial \bom{v}}{\partial t}
      +\frac{\dot{a}}{a}\bom{v}
      +\frac{1}{a}\left(\bom{v}\cdot\nx
                  \right)\bom{v}
    \right)
    =0.
  \end{equation}
\end{subequations}
Eqs.~(\ref{eq:basic1+basic4}) and (\ref{eq:basic2+basic3}) are
the equations for 
the longitudinal and the transverse mode 
with respect to the Eulerian coordinates,
respectively.
We rewrite the system (\ref{sys:basic_sum})
to the Lagrangian equations
by transforming the Eulerian partial derivative into the Lagrangian one
\begin{eqnarray}
\label{def:x derivative}
  \frac{\partial}{\partial x_{i}}
  =
  \frac{\partial q_{j}}{\partial x_{i}}
  \frac{\partial}{\partial q_{j}}
  &=&
  \left(\delta_{ji}-\frac{\partial A_{j}}{\partial x_{i}}
  \right)
  \frac{\partial}{\partial q_{j}}\nonumber\\
  &:=&
  \left(\delta_{ji}+B_{ji}
  \right)
  \frac{\partial}{\partial q_{j}}
\end{eqnarray}
and introducing the Lagrangian time derivative
\begin{equation}
\label{def:Lagrangian time derivative}
  \frac{d}{dt}:=\frac{\partial}{\partial t}
                +\frac{1}{a}\bom{v}\cdot\nx.
\end{equation}
The results are
\begin{subequations}
\label{sys:Lagrangian equations}
  \begin{eqnarray}
  \label{eq:q-longitudinal}
    &&
    \frac{d}{dt}
    \left[a^3 
          \left\{{\cal D}(\nq\cdot\boml{A})
                 -4\pi G \rho_{b}\nq\cdot\boml{A}
          \right\}
          \right]\nonumber\\
    &&\hspace{0.1cm}=\hspace{-1mm}
    -B_{ij}\hspace{-1mm}
    \left[\frac{d}{dt}\left\{a^3\left({\cal D}(A_{j,i})-4\pi G\rho_{b}A_{j,i}
                                \right)
                      \right\}
    \right]\hspace{-1mm}
    +a^3\nx\cdot\left[(\dot{\bom{A}}\cdot\nabla_{x}){\cal D}(\bom{A})
                       -\left(\nabla_{x}\cdot {\cal D}(\bom{A})
                        \right)
                        \dot{\bom{A}}
                \right],\nonumber\\
  \end{eqnarray}
  \begin{equation}
  \label{eq:q-transverse}
    {\cal D}(\epsilon_{ijk}\tp{A}_{k,j})
   =-\epsilon_{ijk}B_{lj}{\cal D}(A_{k,l}),
  \end{equation}
\end{subequations}
where ${\cal D}$ is the differential operator satisfying
\begin{equation}
\label{def:D}
  {\cal D}(F):=\left(\frac{d^2 F}{dt^2}
                     +2\frac{\dot{a}}{a}\frac{dF}{dt}
                     \right),
\end{equation}
$\epsilon_{ijk}$ is Levi-Civita asymmetric tensor,
$\epsilon_{123}:=1$,
and $\nq$ is the Lagrangian nabla operator.
The detail of the second term in right-hand side of (\ref{eq:q-longitudinal})
is in Appendix \ref{Detail of equation for longitudinal mode}.
Eqs.~(\ref{eq:q-longitudinal}) and (\ref{eq:q-transverse})
are the equations for the longitudinal and transverse mode
with respect to the Lagrangian coordinates,
respectively.
The system (\ref{sys:Lagrangian equations}) is the Lagrangian equations
which are necessary for obtaining the density fluctuation.

Next,
we shall consider the expression of the vorticity in the Lagrangian space.
Here,
we define the vorticity as $\bom{w}:=(1/a)\nx\times\bom{v}$.
By transforming $\bom{x}$ into $\bom{q}$,
we obtain
\begin{equation}
\label{eq:vorticity_relation}
  w_{i}=\frac{1}{a}[\nq\times\bom{v}]_{i}
          +\frac{1}{a}\epsilon_{ijk}B_{lj}v_{k,l}.
\end{equation}
However,
there exists another expression of the vorticity.
In the Lagrangian representation,
the equation for the transverse mode with respect to the Eulerian space
(\ref{eq:basic2+basic3}) is solved exactly as
(see Ref.~\citen{Buchert1})
\begin{equation}
\label{sol:cauchy_integral}
  \bom{w}=\frac{1}{a^2}
          (1+\delta)
          \left(\bom{w}(\bom{q},t_{0})\cdot\nq\right)\bom{x}(\bom{q},t),
\end{equation}
where $\bom{w}(\bom{q},t_{0})$ is the primordial vorticity.
Eq.~(\ref{sol:cauchy_integral}) represents
that the vorticity couples to the enhancement of the density fluctuation.
This fact is not respected in the Eulerian theory.
Naturally,
these two expressions of the vorticity 
(\ref{eq:vorticity_relation}) and (\ref{sol:cauchy_integral})
are equivalent.
We stress that
the Lagrangian method enables 
not only to calculate the density fluctuation straightforward
but to clarify vorticity-density relation.

In next section,
we will solve the Lagrangian equations by perturbative method
and obtain the concrete expression 
of the density fluctuation and the vorticity.


\section{Lagrangian perturbative approximation}
\label{sec:Lagrangian perturbative approximation}

In this section,
we solve the Lagrangian equations
(\ref{eq:q-longitudinal}) and (\ref{eq:q-transverse})
by perturbative method up to the third-order
and obtain the determinant $J$.
By calculating $J$,
we obtain the concrete expression of 
the density fluctuation and the vorticity
from Eqs.~(\ref{def:density}) and (\ref{sol:cauchy_integral}),
respectively.
Our treatment is different from
the standard Lagrangian perturbation theory
in considering the vorticity.
Naturally,
when the irrotational condition is assumed,
our results are consistent with the standard Lagrangian perturbation theory.
The reason the perturbative approach is introduced
is that the right-hand sides of the system (\ref{sys:Lagrangian equations})
are non-linear.
However,
since the density fluctuation itself has no restriction,
Lagrangian perturbative approximation enables to
describe the high-density regions.
(The restriction by the Lagrangian perturbation approach is that
 we require the smallness of the deviation of the particle trajectory
 from the uniform Hubble flow.)
In the Lagrangian perturbation approach,
the perturbative quantity is the displacement vector:
\begin{equation}
  \bom{x}(\bom{q},t)
  =\bom{q}+\bom{A}_{(1)}(\bom{q},t)+\bom{A}_{(2)}(\bom{q},t)
   +\bom{A}_{(3)}(\bom{q},t)+\cdots,
\end{equation}
where
$\bom{A}_{(1)}(\bom{q},t)$ corresponds to the first-order approximation,
$\bom{A}_{(2)}(\bom{q},t)$ to the second-order approximation,
and so on.
We derive the perturbative equations for 
$\bom{A}_{(1)}$, $\bom{A}_{(2)}$ and $\bom{A}_{(3)}$
and then solve these equations.
Hereafter,
we call the longitudinal~(transverse) mode
with respect to the Lagrangian coordinates
simply ``the longitudinal~(transverse) mode''.


\subsection{First-order solutions}
\label{sec:First-order solutions}

The first-order solutions for the longitudinal and the transverse mode
are obtained by Buchert.
\cite{Buchert1} \ 

First,
we consider the longitudinal mode.
The equation for the first-order longitudinal mode is
\begin{equation}
\label{eq:1st-longitudinal}
  \frac{d}{dt}
  \left[a^3 
        \left\{{\cal D}(\bomlo{A}{1})
               -4\pi G \rho_{b}\bomlo{A}{1}
        \right\}
  \right]
  =\nq\times\bom{K},
\end{equation}
where
$\bom{K}$ is an arbitrary vector potential.
The existence of the quantity $\bom{K}$ means that
the decomposition such as $\bom{A}=\boml{A}+\bomt{A}$ is not unique;
we may always add to $\boml{A}$
a rotation-free solution of $\nq\cdot\boml{A}=0$.
We can impose additional condition for eliminating this freedom.
So we shall assume $\nq\cdot\bom{K}=0$.
Integrating Eq.~(\ref{eq:1st-longitudinal}),
we obtain 
\begin{equation}
\label{eq:1st-longitudinal2}
  {\cal D}(\bomlo{A}{1})-4\pi G \rho_{b}\bomlo{A}{1}
  =\frac{\bom{C}}{a^3},
  \mbox{ }
  \frac{d\bom{C}}{dt}=0.
\end{equation}
By choosing $\bom{C}=0$,
we obtain
\begin{equation}
\label{sol:1st-longitudinal}
  \bomlo{A}{1}=t^{2/3}\bomlo{X}{1}(\bom{q})
               +t^{-1}\bomlo{X}{1d}(\bom{q}).
\end{equation}
(The Lagrangian perturbative solutions at any order
 separate with respect to $\bom{q}$ and $t$.
 See Ref.~\citen{Ehlers}.)
The terms proportional to $t^{2/3}$ and $t^{-1}$ are called 
the growing mode and the decaying mode,
respectively.

Next,
we consider the transverse mode.
The equation for the first-order transverse mode is
\begin{equation}
\label{eq:1st-transverse}
  {\cal D}(\bomto{A}{1})
  =\nq\varphi,
\end{equation}
where
$\varphi$ is an arbitrary scalar potential.
We assume $\nq\varphi=0$ for eliminating the freedom.
The solution is
\begin{equation}
\label{sol:1st-transverse}
  \bomto{A}{1}=t^{-1/3}\bomto{X}{1}(\bom{q}).  
\end{equation}
We shall call this solution the rotational mode.


\subsection{Second-order solutions}
\label{sec:Second-order solutions}

The general irrotational second-order solutions are obtained
by Buchert and Ehlers
\cite{Buchert2}
in the case of density parameter $\Omega=1$.
The leading terms of the irrotational second-order solution
in the case of arbitrary density parameter are
derived by Bouchet et al..
\cite{Bouchet2} \ 
In order to avoid notational complexity,
we neglect the decaying mode $\bomlo{X}{1d}$ hereafter.

First,
we consider the longitudinal mode.
The equation for the second-order longitudinal mode is
\begin{eqnarray}
\label{eq:2nd-longitudinal}
  &&
  \frac{d}{dt}
  \left[a^3 
        \left\{{\cal D}(\nq\cdot\bomlo{A}{2})
               -4\pi G \rho_{b}\nq\cdot\bomlo{A}{2}
        \right\}
  \right]\nonumber\\
  &&\mbox{ }=
  \hspace{-1mm}
  -B_{(1)ij}
  \left[
        \frac{d}{dt}
        \left\{a^3 
               \left({\cal D}(A_{(1)j,i})
                     -4\pi G \rho_{b}A_{(1)j,i}
               \right)
        \right\}
  \right]\nonumber\\
  &&\hspace{1cm}
  +
  a^3\left[\dot{A}_{(1)i,j}{\cal D}(\lpo{A}{1}{j,i})
            -\dot{A}_{(1)i,i}{\cal D}(\lpo{A}{1}{j,j})
      \right].
\end{eqnarray}
The particular solution is
\begin{equation}
\label{sol:2nd-longitudinal}
  \bomlo{A}{2}=t^{4/3}\bomlo{X}{2a}(\bom{q})
               +t^{-2/3}\bomlo{X}{2b}(\bom{q}),
\end{equation}
where
\begin{equation}
\label{def:X2a}
  \nq\cdot\bomlo{X}{2a}=\frac{3}{14}\left[\lpo{X}{1}{i,j}\lpo{X}{1}{j,i}
                                          -\left(\nq\cdot\bomlo{X}{1}\right)^2
                                    \right],
\end{equation}
\begin{equation}
\label{def:X2b}
  \nq\cdot\bomlo{X}{2b}=\frac{3}{4}\tpo{X}{1}{i,j}\tpo{X}{1}{j,i}.
\end{equation}
Although the solution for Eq.~(\ref{eq:2nd-longitudinal}) consists of
the homogeneous and the particular solution,
we omit the homogeneous solution.
We focus on the behavior of the first-order quantities,
namely,
the growing mode $\bomlo{X}{1}$ and the rotational mode $\bomto{X}{1}$,
in higher-order.
Since
the homogeneous equation is the same form as the first-order equation,
the behavior of the homogeneous solutions is investigated easily.
Now the solution (\ref{sol:2nd-longitudinal}) consists of two parts.
One is $\bomlo{X}{2a}$ described by using two growing modes.
Another is $\bomlo{X}{2b}$ 
described by using two rotational modes.
In the case where the rotational mode is considered,
the solution $\bomlo{A}{2}$ includes the $\bomto{X}{1}$,
although $\bomlo{A}{2}$ describes the longitudinal mode.
Furthermore,
there is no cross term consisting of $\bomlo{X}{1}$ and $\bomto{X}{1}$.
This cross term is included the transverse mode as mentioned below.

Next,
we consider the transverse mode.
The equation for the second-order transverse mode is
\begin{equation}
\label{eq:2nd-transverse}
  {\cal D}(\epsilon_{ijk}\tpo{A}{2}{k,j})
  =
  -\epsilon_{ijk}B_{(1)lj}{\cal D}(A_{(1)k,l}).
\end{equation}
In the same sense as that of the longitudinal mode,
we shall obtain the particular solution for Eq.~(\ref{eq:2nd-transverse}).
The particular solution is
\begin{equation}
\label{sol:2nd-transverse}
  \bomto{A}{2}=t^{1/3}\bomto{X}{2},
\end{equation}
where
\begin{equation}
\label{def:S2}
  \left[\nq\times\bomto{X}{2}\right]_{i}
  =
  -3\epsilon_{ijk}\lpo{X}{1}{j,l}\tpo{X}{1}{l,k}.
\end{equation}
The second-order transverse mode is induced by
the coupling of the rotational mode to the growing mode.
Note that this coupling term is growing. 
On the other hand,
the cross term of two $\bomlo{X}{1}$ or two $\bomto{X}{1}$ does not exist
in Eq.~(\ref{sol:2nd-transverse}).
Therefore,
if the rotational mode is zero,
the second-order transverse mode vanishes.
Actually,
we must consider the fact that the decaying mode couples to
$\bomlo{X}{1}$ or $\bomto{X}{1}$.
However,
the coupling of the decaying mode is 
different from the coupling of $\bomto{X}{1}$ to the $\bomlo{X}{1}$
in decaying at late time.
Although both the decaying mode and the rotational mode decay,
the former is more effective
when they couple to the growing mode in higher-order.


\subsection{Third-order solutions}
\label{sec:Third-order solutions}

The irrotational third-order solution is obtained,
for example,
by Buchert
\cite{Buchert3},
Bouchet et al.,
\cite{Bouchet3}
and Catelan
\cite{Paolo}.

First,
we consider the longitudinal mode.
The equation for the third-order longitudinal mode is
\begin{eqnarray}
\label{eq:3rd-longitudinal}
  &&
  \frac{d}{dt}\left[a^3\left\{{\cal D}(\nq\cdot\bomlo{A}{3})
                              -4\pi G\rho_{b}\nq\cdot\bomlo{A}{3}
                       \right\}
              \right]\nonumber\\
  &&\mbox{ }=
  -B_{(1)ij}\frac{d}{dt}\left[a^3\left\{{\cal D}(A_{(2)j,i})
                                        -4\pi G\rho_{b}A_{(2)j,i}
                                 \right\}
                        \right]\nonumber\\
  &&\hspace{1cm}
  -B_{(2)ij}\frac{d}{dt}\left[a^3\left\{{\cal D}(A_{(1)j,i})
                                        -4\pi G\rho_{b}A_{(1)j,i}
                                 \right\}
                        \right]\nonumber\\
  &&\hspace{1cm}
  +a^3\left[\dot{A}_{(1)i,j}{\cal D}(A_{(2)j,i})
            +\dot{A}_{(2)i,j}{\cal D}(A_{(1)j,i})
      \right.
      \left.
            -\dlpo{A}{1}{i,i}{\cal D}(\lpo{A}{2}{j,j})
            -\dlpo{A}{2}{i,i}{\cal D}(\lpo{A}{1}{j,j})
      \right]\nonumber\\
  &&\hspace{1cm}
  +a^3\left[B_{(1)ij}\dot{A}_{(1)j,k}{\cal D}(A_{(1)k,i})
            +B_{(1)ij}{\cal D}(A_{(1)j,k})\dot{A}_{(1)k,i}
      \right.\nonumber\\
  &&\hspace{3cm}
      \left.
            -B_{(1)ij}{\cal D}(A_{(1)j,i})\dlpo{A}{1}{k,k}
            -B_{(1)ij}\dot{A}_{(1)j,i}{\cal D}(\lpo{A}{1}{k,k})
      \right].
\end{eqnarray}
The particular solution is
\begin{equation}
\label{sol:3rd-longitudinal}
  \bomlo{A}{3}=t^2\bomlo{X}{3a}+t\bomlo{X}{3b}
               +\frac{2}{75}t^{-1}(3+5\ln t)\bomlo{X}{3c},
\end{equation}
where
\begin{equation}
\label{def:3a_1}
  \nq\cdot\bomlo{X}{3a}=\frac{5}{9}\left[\lpo{X}{1}{i,j}\lpo{X}{2a}{j,i}
                                          -\lpo{X}{1}{i,i}\lpo{X}{2a}{j,j}
                                   \right]
                        -\frac{1}{3}\det(\lpo{X}{1}{i,j}),
\end{equation}
\begin{eqnarray}
  \nq\cdot\bomlo{X}{3b}&=&
  \frac{1}{3}\lpo{X}{1}{i,j}\tpo{X}{2}{j,i}
  +\frac{7}{3}\lpo{X}{2a}{i,j}\tpo{X}{1}{j,i}\nonumber\\
  &&\mbox{ }
  -\left[\lpo{X}{1}{i,j}\lpo{X}{1}{j,k}\tpo{X}{1}{k,i}
                  -\lpo{X}{1}{i,j}\tpo{X}{1}{j,i}\lpo{X}{1}{k,k}
   \right],
\end{eqnarray}
\begin{equation}
  \nq\cdot\bomlo{X}{3c}=2\lpo{X}{2b}{i,j}\tpo{X}{1}{j,i}
                        -\tpo{X}{1}{i,j}\tpo{X}{1}{j,k}\tpo{X}{1}{k,i}.    
\end{equation}
The third-order solution (\ref{sol:3rd-longitudinal}) does not contain
the term consisting of two $\bomto{X}{1}$ and one $\bomlo{X}{1}$.
Furthermore,
note that
the third-order solution includes the term proportional to $t^{-1}\ln t$,
although the solutions up to the second-order are described
in the form of simple power of~$t$.
Since this term consists of three $\bomto{X}{1}$,
the term is anticipated proportional to $t^{-1}$,
namely,
$(t^{-1/3})^3$.
However,
the solution proportional to $t^{-1}$ is same form 
as the first-order decaying solution,
that is,
the homogeneous solution of the Lagrangian equation.
Therefore,
the term proportional to $t^{-1}\ln t$ is necessary.

Next,
we consider the transverse mode.
The equation for the third-order transverse mode is
\begin{equation}
\label{eq:3rd-transverse}
  {\cal D}(\epsilon_{ijk}\tpo{A}{3}{k,j})
  =-\epsilon_{ijk}\left[B_{(1)lj}{\cal D}(A_{(2)k,l})
                        +B_{(2)lj}{\cal D}(A_{(1)k,l})
                  \right].
\end{equation}
The particular solution is
\begin{equation}
\label{sol:3rd-transverse}
  \bomto{X}{3}=t^2\bomto{X}{3a}+t\bomto{X}{3b}+\ln t \bomto{X}{3c}
               +t^{-1}\bomto{X}{3d},
\end{equation}
where
\begin{equation}
\label{3a}
  \epsilon_{ijk}\tpo{X}{3a}{k,j}=\frac{1}{3}\epsilon_{ijk}
                                 \lpo{X}{1}{l,j}\lpo{X}{2a}{k,l},
\end{equation}
\begin{equation}
  \epsilon_{ijk}\tpo{X}{3b}{k,j}=-\frac{1}{3}\epsilon_{ijk}
                                   \left[\lpo{X}{1}{l,j}\tpo{X}{2}{l,k}
                                         +5\lpo{X}{2a}{l,j}\tpo{X}{1}{l,k}
                                   \right],  
\end{equation}
\begin{equation}
  \epsilon_{ijk}\tpo{X}{3c}{k,j}=\frac{2}{3}\epsilon_{ijk}
                                 \left[\tpo{X}{1}{l,j}\tpo{X}{2}{l,k}
                                       -2\lpo{X}{1}{l,j}\lpo{X}{2b}{k,l}
                                 \right],
\end{equation}
\begin{equation}
\label{3d}
  \epsilon_{ijk}\tpo{X}{3d}{k,j}=\frac{1}{3}\epsilon_{ijk}
                                 \tpo{X}{1}{l,j}\lpo{X}{2b}{k,l}.  
\end{equation}
Note that
even though the irrotational condition is assumed,
the non-zero transverse mode exists
in the first term in the right-hand side of Eq.~(\ref{sol:3rd-transverse}),
as Buchert
\cite{Buchert3}
pointed out.
In deriving (\ref{3a})$\sim$(\ref{3d}),
by using the following relation
\begin{equation}
  \epsilon_{ijk}\lpo{X}{1}{j,m}\left[\lpo{X}{1}{m,l}\tpo{X}{1}{l,k}
                                     -\lpo{X}{1}{k,l}\tpo{X}{1}{l,m}
                               \right]
  =\frac{1}{3}
  \epsilon_{ijk}
    \lpo{X}{1}{j,l}\left[\tpo{X}{2}{l,k}-\tpo{X}{2}{k,l}
                   \right],
\end{equation}
\begin{equation}
  \epsilon_{ijk}\lpo{X}{1}{j,l}\tpo{X}{1}{l,m}\tpo{X}{1}{m,k}
  =\frac{1}{3}\epsilon_{ijk}\tpo{X}{1}{l,j}
               \left[\tpo{X}{2}{l,k}-\tpo{X}{2}{k,l}
               \right],
\end{equation}
we write the triplet of $\bom{X}_{(1)}$ as the term including 
the second-order quantity.
By the same reason in the case of the longitudinal solution,
the term proportional to $\ln t$ exists
in the third-order transverse solution.


\subsection{Relation between vorticity and density fluctuation}

In this subsection,
we calculate the concrete form of the density fluctuation and the vorticity
by using the Lagrangian perturbation formalism we have obtained above.

From the definition of the density fluctuation
Eq.~(\ref{def:density}),
we obtain the density fluctuation as follows:
\begin{eqnarray}
\label{sol:full_density}
  \delta &=& \dfrac{1}{\det\left[\delta_{ij}+A_{(1)i,j}+A_{(2)i,j}+A_{(3)i,j}
                           \right]
                      }
             -1\nonumber\\
         &=& \dfrac{1}{1+J_{(1)}+J_{(2)}+J_{(3)}+\ldots}-1,
\end{eqnarray}
where
\begin{equation}
\label{sol:Jacobian1}
  J_{(1)}=t^{2/3}\nq\cdot\bomlo{X}{1},
\end{equation}
\begin{equation}
\label{sol:Jacobian2}
  J_{(2)}=
     \frac{2}{7}t^{4/3}\left[\left(\nq\cdot\bomlo{X}{1}\right)^2
                             -\lpo{X}{1}{i,j}\lpo{X}{1}{j,i}
                       \right]
     -t^{1/3}\lpo{X}{1}{i,j}\tpo{X}{1}{j,i}
     +\frac{1}{4}t^{-2/3}\tpo{X}{1}{i,j}\tpo{X}{1}{j,i},
\end{equation}
\begin{eqnarray}
\label{sol:Jacobian3}
  J_{(3)}
  &=&
  t^2\left[\frac{4}{9}\left\{\lpo{X}{1}{i,i}\lpo{X}{2a}{j,j}
                             -\lpo{X}{1}{i,j}\lpo{X}{2a}{j,i}
                      \right\}
           +\frac{2}{3}\det(\lpo{X}{1}{i,j})
     \right]\nonumber\\
  &&\mbox{ }
  +t\left[-\frac{2}{3}\lpo{X}{1}{i,j}\tpo{X}{2}{j,i}
          +\frac{4}{3}\lpo{X}{2a}{i,j}\tpo{X}{1}{j,i}
    \right]\nonumber\\
  &&\mbox{ }
  +\left[-\tpo{X}{1}{i,j}\tpo{X}{2}{j,i}
         +\left\{\lpo{X}{1}{i,i}\lpo{X}{2b}{j,j}
                 -\lpo{X}{1}{i,j}\lpo{X}{2b}{j,i}
         \right\}
  \right.\nonumber\\
  &&\hspace{2cm}
  \left.+\lpo{X}{1}{i,j}\tpo{X}{1}{j,k}\tpo{X}{1}{k,i}
         -\frac{1}{2}\tpo{X}{1}{i,j}\tpo{X}{1}{j,i}\lpo{X}{1}{l,l}
  \right]\nonumber\\
  &&\mbox{ }
  +\frac{1}{75}t^{-1}(20\ln t -63)\lpo{X}{2b}{i,j}\tpo{X}{1}{j,i}\nonumber\\
  &&\mbox{ }
  +\frac{1}{75}t^{-1}(19-10\ln t)\tpo{X}{1}{i,j}\tpo{X}{1}{j,k}\tpo{X}{1}{k,i}.
\end{eqnarray}
Although we represent $J$ only up to the third-order here,
we can calculate $J$ up to the ninth-order.
Recall that 
it is unnecessary that the density fluctuation is expanded perturbatively
in spite of $J$ being the perturbative quantity.
By linearizing Eq.~(\ref{sol:full_density}),
we recover the Eulerian linear behavior
presenting that the vorticity decouples from the density enhancement.
From the quantities $J_{(2)}$ and $J_{(3)}$,
we see that the coupling of 
the rotational mode $\bomto{X}{1}$ to the growing mode $\bomlo{X}{1}$
plays an important role in the enhancement of the density fluctuation.
Because this coupling is growing.
The decaying mode we have omitted also can
contribute to the enhancement of the density fluctuation.
If we write the term containing one $\bomlo{X}{1d}$
and two $\bomlo{X}{1}$ in $J_{(3)}$,
this term is proportional to $t^{1/3}$,
namely,
$t^{-1}(t^{2/3})^2$.
However,
since the term coupling of the rotational mode is proportional to $t$,
the rotational mode makes more important contribution
to the enhancement of the density fluctuation
than the decaying mode.
The importance of the coupling of the rotational mode to the growing mode
is also seen 
from vorticity-density relation (see Eq.~(\ref{sol:cauchy_integral}))
\begin{equation}
\label{eq:vd-relation}
  \bom{w}=\frac{1}{a^2}
          \dfrac{\left(\bom{w}(\bom{q},t_{0})\cdot\nq\right)
                 (\bom{q}+\bom{A}_{(1)}+\bom{A}_{(2)}+\bom{A}_{(3)})}
                {1+J_{(1)}+J_{(2)}+J_{(3)}+\ldots}.
\end{equation}
The vorticity-density relation shows that
the behavior of the vorticity depends on
both the density fluctuation ($\delta$ or $J$)
and the deformation of the fluid ($x_{i,j}$ or $A_{i,j}$).
For realizing the concrete form of the vorticity,
we shall expand Eq.~(\ref{eq:vd-relation}) perturbatively.
(Note that the vorticity is actually the non-perturbative quantity
 in the Lagrangian form.)
We obtain
\label{sol:vorticity}
\begin{equation}
\label{sol:vorticity1}
  \bom{w}_{(1)}=\frac{1}{a^2}\bom{w}_{(1)}(\bom{q}),
\end{equation}
\begin{eqnarray}
\label{sol:vorticity2}
  \bom{w}_{(2)}&=&
                \frac{1}{a^2}
                \left[t^{2/3}\left\{\left(\bom{w}_{(1)}(\bom{q})\cdot\nq\right)
                                    \bomlo{X}{1}
                                    -\bom{w}_{(1)}(\bom{q})
                                     \left(\nq\cdot\bomlo{X}{1}\right)
                             \right\}
                \right.\nonumber\\
               &&\hspace{1cm}
                \left.
                     +t^{-1/3}\left(\bom{w}_{(1)}(\bom{q})\cdot\nq\right)
                              \bomto{X}{1}
                \right],
\end{eqnarray}
\begin{eqnarray}
\label{sol:vorticity3}
  \bom{w}_{(3)}
  &=&
  \frac{1}{a^2}\left[t^{4/3}\left\{\frac{1}{7}\bom{w}_{(1)}(\bom{q})
                                   \left(5\left(\nq\cdot\bomlo{X}{1}
                                          \right)^2
                                         +2\lpo{X}{1}{a,b}\lpo{X}{1}{b,a}
                                   \right)
               \right.\right.\nonumber\\
  &&\hspace{2.5cm}\left.
                                   -\left(\bom{w}_{(1)}(\bom{q})\cdot\nq
                                    \right)\bomlo{X}{1}
                                    \left(\nq\cdot\bomlo{X}{1}
                                    \right)
                                   +\left(\bom{w}_{(1)}(\bom{q})\cdot\nq
                                    \right)\bomlo{X}{2a}
                           \right\}\nonumber\\
   &&\hspace{0.8cm}
   +t^{1/3}\left\{\bom{w}_{(1)}(\bom{q})\lpo{X}{1}{a,b}\tpo{X}{1}{b,a}
                  -\left(\bom{w}_{(1)}(\bom{q})\cdot\nq\right)
                   \bomto{X}{1}
                   \left(\nq\cdot\bomlo{X}{1}\right)
           \right.\nonumber\\
   &&\hspace{2.5cm}\left.
                  +\left(\bom{w}_{(1)}(\bom{q})\cdot\nq\right)\bomto{X}{2}
           \right\}\nonumber\\
   &&\hspace{0.8cm}\left.
   +t^{-2/3}\left\{-\frac{1}{4}\bom{w}_{(1)}(\bom{q})\tpo{X}{1}{a,b}
                                                     \tpo{X}{1}{b,a}
                   +\left(\bom{w}_{(1)}(\bom{q})\cdot\nq\right)\bomlo{X}{2b}
            \right\}
            \right].
\end{eqnarray}
Here,
we assume 
%
\begin{equation}
  \bom{w}(\bom{q},t_0):=\bom{w}_{(1)}(\bom{q})
                       =-\frac{1}{3}\nq\times\bomto{X}{1}.
\end{equation}
Same results are also derived from Eq.~(\ref{eq:vorticity_relation}),
that is,
the transformation of
the Eulerian coordinates to the Lagrangian one
(see Appendix \ref{sec:Vorticity-density relation}).
The results (\ref{sol:vorticity1}) is the same as
the behavior of the vorticity in Eulerian description.
Furthermore,
from the results (\ref{sol:vorticity2}) and (\ref{sol:vorticity3}),
we see that
the amplitude of induced vorticity becomes strong
by the coupling of the rotational mode (initial vorticity)
to many growing modes.
This shows that
the primordial vorticity is effective in the high-density regions.
The exact calculation of the vorticity Eq.~(\ref{eq:vd-relation})
will confirm this fact.


\section{Summary and conclusion}
\label{Summary and conclusions}

In this paper,
we have investigated the relation
between the vorticity and the density fluctuation
by preparing the rotational Lagrangian perturbation theory.
The fact that the vorticity couples to the density enhancement
is already discussed by Buchert.
\cite{Buchert1} \ 
However,
we obtain more concrete vorticity-density relation
by solving the perturbative equations
up to the third-order
for both longitudinal and transverse displacement vector.
The coupling of the transverse mode to the growing mode
enables the vorticity 
to contribute to the enhancement of the density fluctuation.
Furthermore,
growing vortical modes are induced by
the coupling of initial vorticity to the growing mode.
These results show the fact that
the primordial vorticity is not negligible in high-density regions.
Therefore,
we must reconsider 
the importance of the primordial vorticity effect on the density fluctuation.
Now we also have pointed out the effects of the decaying mode.
It is possible that the first-order decaying mode contributes to
the enhancement of the density fluctuation
by coupling to the growing mode.
However,
since the transverse mode is proportional to $t^{-1/3}$
and the decaying mode to $t^{-1}$,
the former makes more important role 
in the enhancement of the density fluctuation.   

The results we obtained in this paper give a powerful tool
to treat the high-density regions
and behavior of the vorticity.
It is of interest to investigate quantitatively
how the vorticity affects the non-linear evolution of the density fluctuations
in the universe.
It will be the subject of future investigation.


\section*{Acknowledgements}

We would like to thank T.~Kataoka and K.~Konno
for very stimulating suggestions and helpful discussions.
We would also like to thank T.~Buchert
for critical reading of the manuscript and many valuable remarks.


\appendix


\section{Detail of Lagrangian equation for longitudinal mode}
\label{Detail of equation for longitudinal mode}

For the perturbative calculation,
we give the detail of the second term of the right-hand side of
the longitudinal equation (\ref{eq:q-longitudinal}) as follows:
\begin{eqnarray}
  &&
  \nx\cdot\left[\left(\dot{\bom{A}}\cdot\nx\right){\cal D}(\bom{A})
                -\left(\nx\cdot{\cal D}(\bom{A})\right)\dot{\bom{A}}
          \right]\nonumber\\
  &&\mbox{ }=
  \dot{A}_{i,j}{\cal D}(A_{j,i})
  -\dot{A}_{i,i}{\cal D}(A_{j,j})\nonumber\\
  &&\hspace{1cm}
  +B_{ij}\dot{A}_{j,k}{\cal D}(A_{k,i})
  +B_{ij}{\cal D}(A_{j,k})\dot{A}_{k,i}
  -B_{ij}\dot{A}_{j,i}{\cal D}(A_{l,l})
  -B_{ij}{\cal D}(A_{j,i})\dot{A}_{l,l}\nonumber\\
  &&\hspace{1cm}
  +B_{ij,k}\dot{A}_{j}{\cal D}(A_{k,i})
  -B_{ij,k}\dot{A}_{k}{\cal D}(A_{j,i})\nonumber\\
  &&\hspace{1cm}
  +B_{ij}\left[B_{kl}\dot{A}_{l}{\cal D}(A_{j,k})
               -B_{kl}{\cal D}(A_{l,k})\dot{A}_{j}
         \right]_{,i}.
\end{eqnarray}
%


\section{Calculation of $B_{ij}$}
\label{Calculation of B}

When we solve the perturbative equations,
the quantities $B_{(1)ij}$ and $B_{(2)ij}$ are necessary.
So we present the derivation of their quantities here.

The quantity $B_{ij}$ is defined by using the cofactor $C_{ij}$
as follows:
\begin{equation}
\label{def:BC}
  B_{ij}=\sum_{n=1}\left(J^{-1}C_{ji}\right)_{(n)},
  \mbox{ }
  C_{ji}=\frac{1}{2}\epsilon_{jab}\epsilon_{icd}x_{a,c}x_{b,d}.
\end{equation}
Using Eq.~(\ref{def:BC}),
we obtain
\begin{equation}
  B_{(1)ij}=-A_{(1)i,j},
\end{equation}
\begin{equation}
  B_{(2)ij}=A_{(1)i,k}A_{(1)k,j}-A_{(2)i,j}.
\end{equation}
%


\section{Useful relations to obtain the expression of the vorticity}
\label{sec:Vorticity-density relation}

We have presented the expression of the vorticity
in Eqs.~(\ref{sol:vorticity1}), (\ref{sol:vorticity2}) 
and (\ref{sol:vorticity3}).
These results are derived by vorticity-density relation
Eq.~(\ref{sol:cauchy_integral}).
We also obtain the same results from Eq.~(\ref{eq:vorticity_relation}),
that is,
the transformation of
the Eulerian coordinates to the Lagrangian one.
We list the useful relations to obtain the expression of the vorticity 
from Eq.~(\ref{eq:vorticity_relation}) as follows:
\begin{equation}
  \epsilon_{lmn}\tpo{X}{1}{n,m}\tpo{X}{1}{i,l}
 =\epsilon_{ijk}\tpo{X}{1}{j,l}\tpo{X}{1}{l,k},
\end{equation}
\begin{equation}
  \epsilon_{lmn}\tpo{X}{1}{n,m}\lpo{X}{1}{i,l}
 =\epsilon_{ijk}\left[\tpo{X}{1}{k,j}\left(\nq\cdot\bomlo{X}{1}\right)
                      +\lpo{X}{1}{j,l}\left(\tpo{X}{1}{l,k}-\tpo{X}{1}{k,l}
                                      \right)
                \right],
\end{equation}
\begin{equation}
  \epsilon_{ijk}\tpo{X}{1}{k,j}\tpo{X}{1}{l,m}\tpo{X}{1}{m,l}
 =-2\epsilon_{ijk}\tpo{X}{1}{j,l}\tpo{X}{1}{l,m}\tpo{X}{1}{m,k}, 
\end{equation}
\begin{equation}
  \epsilon_{lmn}\tpo{X}{1}{n,m}\tpo{X}{2}{i,l}
 =\epsilon_{ijk}\left[\tpo{X}{1}{j,l}-\tpo{X}{1}{l,j}\right]
                \tpo{X}{2}{l,k},
\end{equation}
\begin{eqnarray}
  &&
  \epsilon_{ijk}\tpo{X}{1}{l,j}\left[\tpo{X}{2}{l,k}-\tpo{X}{2}{k,l}
                               \right]\nonumber\\
  &&\mbox{ }=
  3\epsilon_{ljk}\lpo{X}{1}{l,m}\tpo{X}{1}{i,j}\tpo{X}{1}{m,k}\nonumber\\
  &&\mbox{ }=
  3\epsilon_{ijk}\lpo{X}{1}{j,l}\tpo{X}{1}{l,m}\tpo{X}{1}{m,k}\nonumber\\
  &&\mbox{ }=
  3\epsilon_{ijk}\left[\lpo{X}{1}{j,m}\tpo{X}{1}{l,m}\tpo{X}{1}{k,l}
                       -\tpo{X}{1}{k,j}\lpo{X}{1}{l,m}\tpo{X}{1}{m,l}
                       +\lpo{X}{1}{l,m}\tpo{X}{1}{m,j}\tpo{X}{1}{k,l}
                 \right]\nonumber\\
  &&\hspace{1cm}
  +3\epsilon_{lmn}\tpo{X}{1}{n,m}\tpo{X}{1}{i,l}\left(\nq\cdot\bomlo{X}{1}
                                                \right),
\end{eqnarray}
\begin{eqnarray}
  &&
  \epsilon_{ijk}\lpo{X}{1}{j,l}\left[\tpo{X}{2}{l,k}-\tpo{X}{1}{k,l}
                               \right]\nonumber\\
  &&\mbox{ }=
  3\epsilon_{ijk}\lpo{X}{1}{j,l}\left[\lpo{X}{1}{l,m}\tpo{X}{1}{m,k}
                                      -\lpo{X}{1}{k,m}\tpo{X}{1}{m,l}
                                \right]\nonumber\\
  &&\mbox{ }=
  3\left[\epsilon_{lmn}\tpo{X}{1}{n,m}\lpo{X}{1}{i,l}
         \left(\nq\cdot\bomlo{X}{1}\right)
         +\epsilon_{ijk}\lpo{X}{1}{j,l}\lpo{X}{1}{l,m}\tpo{X}{1}{k,m}
    \right]\nonumber\\
  &&\hspace{1cm}
  -\frac{3}{2}\epsilon_{ijk}\tpo{X}{1}{k,j}
  \left[\left(\nq\cdot\bomlo{X}{1}\right)^2
        +\lpo{X}{1}{l,m}\lpo{X}{1}{m,l}
  \right].
\end{eqnarray}




\begin{thebibliography}{99}
\bibitem{Peebles}
  P.~J.~E.~Peebles,
  {\it The Large Scale Structure of the Universe}
  (Princeton Univ. Press, Princeton, 1980).
\bibitem{Doroshkevich}
  A.~G.~Doroshkevich,
  Astrophys.~Lett.~{\bf 14}~(1973),~11.
\bibitem{Hoyle}
  F.~Hoyle, in {\it Problems of Cosmical Aerodynamics},
  ed.~J.~M.~Burgers and H.~C.~van de Hulst
  (Central Air Documents Office, Dayton, 1949), p.195.
\bibitem{Buchert1}
  T.~Buchert, 
  Mon.~Not.~R.~Astron.~Soc.~{\bf 254}~(1992),~729.
\bibitem{Barrow}
  J.~D.~Barrow and P.~Saich,
  Class.~Quantum Grav.~{\bf 10}~(1993),~79.
\bibitem{Ehlers}
  J.~Ehlers and T.~Buchert,
  astro-ph/9609036.
\bibitem{Gotz}
  T.~Buchert and G.~G\"{o}tz,
  J.~Math.~Phys.~{\bf28}~(1987), 2714.
\bibitem{Kasai}
  M.~Kasai, 
  Phys.~Rev.~{\bf D52}~(1995), 5605.
\bibitem{Buchert2}
  T.~Buchert and J.~Ehlers,
  Mon.~Not.~R.~Astron.~Soc.~{\bf 264} (1993), 375.
\bibitem{Bouchet2}
  F.~R.~Bouchet, R.~Juszkiewicz, S.~Colombi and R.~Pellat,
  Astrophys.~J.~{\bf 394} (1992), L5.
\bibitem{Buchert3}
  T.~Buchert,
  Mon.~Not.~R.~Astron.~Soc.~{\bf 267}~(1994),~811. 
\bibitem{Bouchet3}
  F.~R.~Bouchet, S.~Colombi, E.~Hivon and R.~Juszkiewicz,
  Astron.~Astrophys. {\bf 296} (1995),~575.
\bibitem{Paolo}
  P.~Catelan,
  Mon.~Not.~R.~Astron.~{\bf 276}~(1995),~115.
\end{thebibliography}
\end{document}